# Determining phase-space properties of the IHEP RFQ output beam using the RMS beam widths from wire-scanners

PENG Jun(彭军)[1,2)]   HUANG Tao(黄涛)   LIU Hua-Chang(刘华昌)   JIANG Hong-Ping(蒋洪平)   LI Peng(李鹏)   LI Fang(李芳)   LI Jian(李健)   LIU Mei-Fei(刘美飞)   MU Zhen-Cheng(慕振成)   MENG Cai(孟才)   MENG Ming(孟鸣)   OUYANG Hua-Fu(欧阳华甫)   RONG Lin-Yan(荣林艳)   TIAN Jian-Min(田建民)   WANG Biao(王标)   WANG Bo(王博)   XU Tao-Guang(徐韬光)   XU Xin-An(徐新安)   YAO Yuan(姚远)   XIN Wen-Qu(辛文曲)   ZHAO Fu-Xiang(赵富祥)   ZENG Lei(曾磊)   ZHOU Wen-Zhong(周文中)

[1] China Spallation Neutron Source, Institute of High Energy Physics, Chinese Academy of Sciences, Dongguan 523803, China

[2] Dongguan Institute of Neutron Science, Dongguan 523808, China

**Abstract:** A beam line is built after the IHEP RFQ for halo study. To determine transverse emittance and ellipse parameters of the RFQ output beam, beam size data obtained from the first two of 14 wire scanners are employed. By using the transfer matrix method and the least square method, a set of linear equations were set up and solved. The solutions were then applied as initial beam parameters in multi-particle simulations to check the method of calculation. It is shown that difference between the simulated RMS beam size and the measured one at the measurement location is less than 7%, which is acceptable in our experiments.

## 1 Introduction

A beam line is installed after the IHEP RFQ, of which the output energy is 3.5MeV and the operation frequency is 352MHz. The purpose of this beam line is to study beam halo formation and to do comparison between measurements and multi-particle simulations. The layout of the beam line is shown in Fig.1 [1]. It consists of 28 quadrupoles. The first four quadrupoles can be independently adjusted to do matching or mismatching. The other 24 quadrupoles are used to form FD lattice. Several types of beam diagnostic devices are installed. The key diagnostic component is a group of 14 wire scanners, which locate in the middle of drift spaces between quadrupoles. A pair of wire scanners which locate after the 5$^{th}$ and 6$^{th}$ quadrupoles are used to determine horizontal and vertical ellipse parameters of the RFQ output beam respectively. These parameters are essential for the upcoming halo experiment. The wire scanners after quadrupole 17 to 22 are used to measure beam profiles in X direction and Y direction alternately. Simultaneously, the wire scanners after quadrupole 23 to 28 are used to measure beam profiles in X direction only.

*Supported by NSFC (91126003) and the State Key Development Program of Basic Research of China (2007CB209904)
1)E-mail: pengjun@mail.ihep.ac.cn

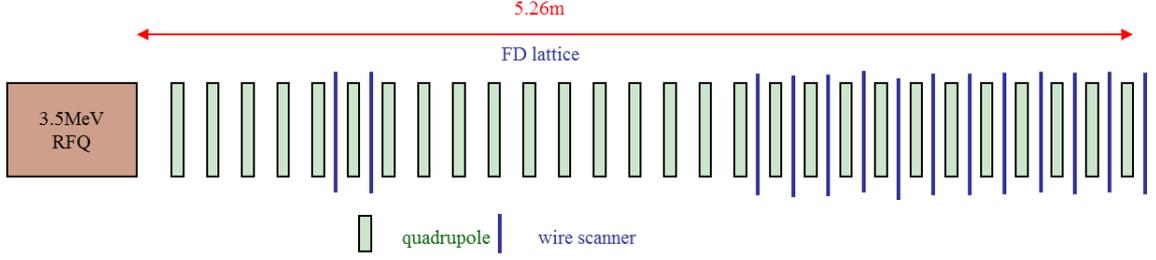

Fig.1. Layout of the beam halo experiment transport line

## 2 Least-squares analysis

For measured beam profile, we generally fitted it with a Gaussian function and then obtained RMS beam size $X_{RMS}$. For a beam transporting through a lattice, the beam ellipse transformation can be expressed in matrix form as [2]

$$\sigma_2 = R\sigma_1 R^T \qquad (1)$$

Where $R$ is the transfer matrix between two positions and $R^T$ is the transpose of the matrix $R$. The $\sigma$-matrix is defined as

$$\sigma = \begin{pmatrix} \beta\varepsilon & -\alpha\varepsilon \\ -\alpha\varepsilon & \gamma\varepsilon \end{pmatrix} \qquad (2)$$

Where $\alpha$, $\beta$, $\gamma$ are the Twiss parameters and $\varepsilon$ is the un-normalized beam emittance. By carrying out matrix multiplication of Eq. (1), we can get

$$\beta_2\varepsilon_2 = R_{11}^2\beta_1\varepsilon_1 - 2R_{11}R_{12}\alpha_1\varepsilon_1 + R_{12}^2\gamma_1\varepsilon_1 \qquad (3)$$

Here, we use $\alpha_1$, $\beta_1$, $\gamma_1$, $\varepsilon_1$ to denote the initial ellipse parameters at the exit of RFQ, and $\alpha_2$, $\beta_2$, $\gamma_2$, $\varepsilon_2$ to denote the ellipse parameters at the measurement location. Since the Twiss parameters satisfy the relationship of $\beta\gamma - \alpha^2 = 1$ and the RMS beam size is equal to $\sqrt{\beta_2\varepsilon_{2,RMS}}$, then Eq. (3) can be expressed as

$$X_{RMS}^2 = R_{11}^2 a + 2R_{11}R_{12}b + R_{12}^2 c \qquad (4)$$

where $a = \beta_1\varepsilon_1$, $b = -\alpha_1\varepsilon_1$, $c = (1+\alpha_1^2)\varepsilon_1/\beta_1$.

According to Eq. (4), if the gradients of quadrupoles between two points are changed, the RMS beam size at the wire scanner will also change. And hence three unknown elements in Eq. (4) can be deduced from a set of equations obtained from three different quadrupole settings.

However, in the experiments, because of errors, the equation sets usually have no solution or the solutions differ greatly between each other. To avoid these problems, more than three independent linear equations are taken into account and the least square method is adopted.

For example, if several groups of beam sizes and quadrupole gradients are known, a set of linear equations can be obtained as:

$$\begin{cases} R_{1,11}^2 a + 2R_{1,11}R_{1,12}b + R_{1,12}^2 c = X_{1,RMS}^2 \\ \quad \cdots \\ R_{n,11}^2 a + 2R_{n,11}R_{n,12}b + R_{n,12}^2 c = X_{n,RMS}^2 \end{cases} \quad (5)$$

where $n \geq 3$.

The variance, or the sum of squared residuals, is calculated as

$$\chi^2 = \sum_{i=1}^{n} (R_{i,11}^2 a + 2R_{i,11}R_{i,12}b + R_{i,12}^2 c - X_{i,RMS}^2)^2 \quad (6)$$

According to the method of least squares, the best solution of Eq.(5) has the property that make $\chi^2$ minimum, which means, at this point, the differential of $\chi^2$ on $a$, $b$, $c$ are equal to zero respectively, as

$$\begin{cases} \dfrac{\partial \sum_{i=1}^{n}(R_{i,11}^2 a + 2R_{i,11}R_{i,12}b + R_{i,12}^2 c - X_{i,RMS}^2)^2}{\partial a} = 0 \\ \dfrac{\partial \sum_{i=1}^{n}(R_{i,11}^2 a + 2R_{i,11}R_{i,12}b + R_{i,12}^2 c - X_{i,RMS}^2)^2}{\partial b} = 0 \\ \dfrac{\partial \sum_{i=1}^{n}(R_{i,11}^2 a + 2R_{i,11}R_{i,12}b + R_{i,12}^2 c - X_{i,RMS}^2)^2}{\partial c} = 0 \end{cases} \quad (6)$$

Expanding these differentials, we get

$$\begin{cases} \sum_{i=1}^{n} 2R_{i,11}^4 a + 4R_{i,11}^3 R_{i,12} b + 2R_{i,12}^2 R_{i,11}^2 c - 2X_{i,RMS}^2 R_{i,11}^2 = 0 \\ \sum_{i=1}^{n} 4R_{i,11}^3 R_{i,12} a + 8R_{i,11}^2 R_{i,12}^2 b + 4R_{i,12}^3 R_{i,11} c - 4X_{i,RMS}^2 R_{i,11} R_{i,12} = 0 \\ \sum_{i=1}^{n} 2R_{i,11}^2 R_{i,12}^2 a + 4R_{i,11} R_{i,12}^3 b + 2R_{i,12}^4 c - 2X_{i,RMS}^2 R_{i,12}^2 = 0 \end{cases} \quad (7)$$

then $a$, $b$, $c$ can be solved from Eq.(7), and the beam parameters will be deduced as

$$\varepsilon = \sqrt{ac - b^2}, \beta = \frac{a}{\sqrt{ac - b^2}}, \alpha = -\frac{b}{\sqrt{ac - b^2}}.$$

## 3 EXPERIMENT DATA ANYLYSIS

As mentioned in the first section, the 1$^{st}$ and the 2$^{nd}$ wire scanner were used to calculate X direction and Y direction beam parameters respectively. For each direction, only one quadrupole between the RFQ and the wire scanner was adjusted to produce different beam size at the wire scanner, while the gradients of other quadrupoles keep constant.

We used a code TRACE-3D [3] to set a model for computing transfer matrices and tracking beam. Because transfer matrices had relations with initial ellipse parameters while beam current wasn't equal to zero, an iteration procedure was used. Firstly, we assumed the beam current was zero and it was easy to get transfer matrices and set up equations. Secondly, we used the zero-current solution as initial ellipse parameters in model to get new transfer matrices and set up new equations. The process was iterated again and again until the difference between two successive solutions was less than $10^{-3}$. In order to get solutions efficiently and correctly, a code was written to solve the equation set, to edit TRACE-3D input file, to call TRACE-3D code and to read the transfer matrix element automatically. In Table 1, the calculated beam parameter is shown. We got two groups of beam parameters for different beam current. It is shown that the beam parameters change slightly when the beam current is increased. In the Y direction, the calculated beam parameters are more close to the PARMTEQM prediction than that in the X direction. And in both directions, there are about 50% difference between the simulated emittance and the calculated ones.

Table 1 Beam parameters at the exit of RFQ

| RFQ exit beam | $\alpha_x$ | $\beta_x$ (mm/mrad) | $\alpha_y$ | $\beta_y$ (mm/mrad) | $\varepsilon_x$ ($\pi$.mm.mrad) | $\varepsilon_y$ ($\pi$.mm.mrad) |
|---|---|---|---|---|---|---|
| PARMTEQM(I=20mA) | -0.126 | 0.079 | -0.709 | 0.218 | 0.205 | 0.21 |
| Experiment-1(I=22mA) | 3.129 | 0.385 | -0.546 | 0.112 | 0.344 | 0.327 |
| Experiment-2(I=26mA) | 3.753 | 0.461 | -0.611 | 0.113 | 0.333 | 0.33 |

Using these calculated beam parameters as initial beam parameters, we simulated proton beam transporting through the beam line and got simulated RMS beam sizes at the measurement location. These simulated beam sizes were compared with those measured ones to check the method of calculation. Besides Trace-3D code, another multi-particle code PARMILA [4] was also used since it can simulate nonlinear space-charge force. In PARMILA simulations, the initial beam distribution was 6D Waterbag distribution, and 50000 macro-particles per bunch were used. Space charge effect was calculated via the 2-dimensional particle in cell (PIC) method. The simulated RMS beam sizes from two codes are shown in Fig. 2 and 3. In Fig. 2, the difference between the measured data and the simulated data is less than 3% in X direction and less than 7% in Y direction. In Fig.3, the difference between the measured data and the simulated data is less than 7% in X direction and less than 3% in Y direction.

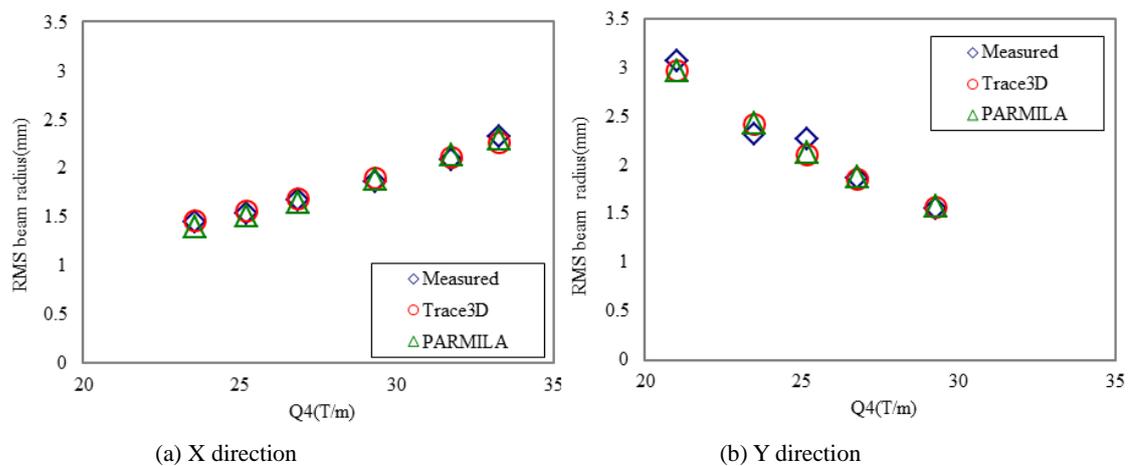

(a) X direction  (b) Y direction

Fig.2. I=22mA, RMS beam size at the measurement location. (The diamonds show the measured data, the circles show the simulated data by using Trace-3D and the triangles show the simulated data by using PARMILA)

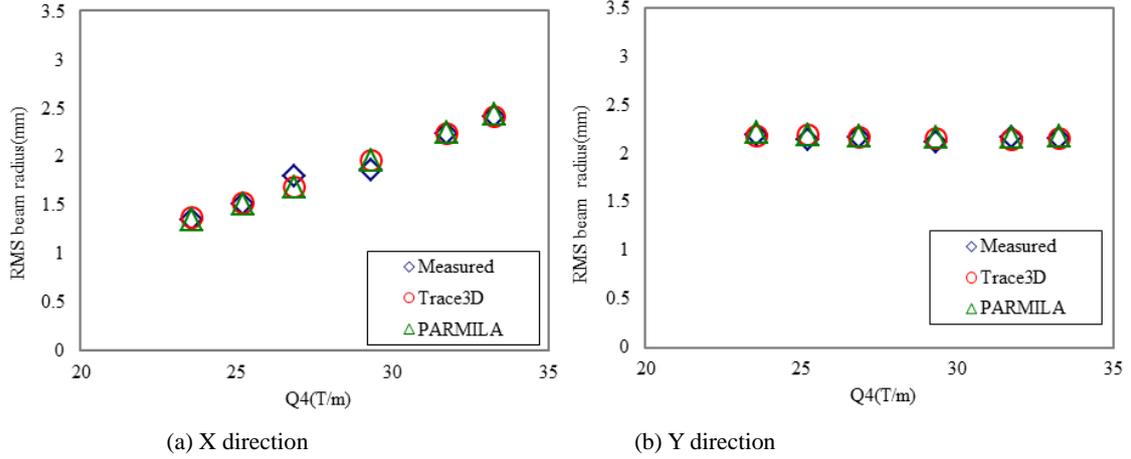

(a) X direction  (b) Y direction

Fig.3. I=26mA, RMS beam size at the measurement location. (The diamonds show the measured data, the circles show the simulated data by using Trace-3D and the triangles show the simulated data by using PARMILA)

## 4 Transverse Matching

Once the initial Twiss parameters were obtained, we used the Trace-3D code to find a proper quad setting for beam matching. However, we found that the matching was not good as proposed. As shown in Fig. 4, we use black dots to represent the measured RMS beam size while quadrupoles are set at the Trace-3D value. As can be seen, the RMS beam sizes measured at the same point of FD periods were not agreed with each other. On the other hand, the measured beam sizes differed from the simulated ones, which were obtained from PARMILA code (The model setting is same with the actual setting). The reasons led to this problem were the measurement uncertainties, the simplified space charge effects in Trace-3D code and the difference between model and real machine. To get a better matching, quadrupole scans were done from 0.8 to 1.2 times of the Trace-3D value. The figure of merit was to obtain equal RMS beam sizes along the beam line. Finally, we found a better matching, of which the gradient of Q4 was 10% higher than the Trace-3D solution. As shown in Fig. 4, the blue circle represents the measured value after quadrupole scans. It is also shown that the measured beam sizes are almost the same with the simulated ones at wire scanners.

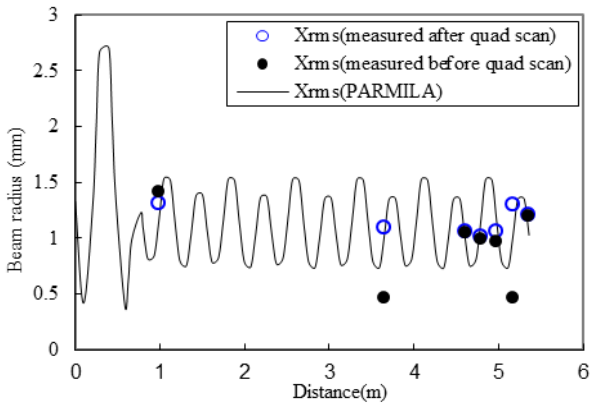

Fig.4. Matched beam envelope along the beam line

## 5 Conclusions

In this paper, we described how to calculate ellipse parameters of the IHEP RFQ output beam using only the RMS beam sizes from wire scanners. However, this method highly depends on the beam quality. For getting accurate parameters, we need to collect a lot of data and abandon ones deviating a lot from others. Once the beam parameters obtained, we can use them to do matching. Due to some unavoidable errors, the matching may be not perfect as proposed. However, it can be refined by adjusting quadrupoles a little based on the Trace-3D solution.

# 用束流 RMS 半径确定 IHEP RFQ 出口处束流参数

彭军 [1,2)]　黄涛　刘华昌　蒋洪平　李鹏　李芳　李健　刘美飞　　慕振成　孟才
孟鸣　欧阳华甫　荣林艳　田建民　王标　王博　徐韬光　徐新安　姚远　辛文曲
赵富祥　曾磊　周文中

[1] 中国科学院高能物理研究所东莞分部，东莞 523803
[2] 东莞中子科学中心，东莞 523808

**摘要**　在 IHEP 的 RFQ 后面，我们建造了一条束运线用于强流质子束的束晕实验研究。如何精确的计算出 RFQ 出口处的束流相椭圆参数是实验中关键的一步。在本文中，我们利用束运线上离 RFQ 最近的两个线扫描器上测得的束流 RMS 半径和变四极磁铁梯度的方法来建立方程组求解，同时，由于测量误差的存在，我们又引入了最小二乘法。对于求出的束流参数，我们将其应用到模拟中，模拟束流通过束运线，得到线扫描器处，模拟的束流 RMS 半径。将其与实测的半径比较，发现误差小于 7%，误差在可接受范围内，证明了算法的可靠性。

**关键词**　强流，质子，线扫描器，椭圆参数, 横向匹配